\documentclass{moriond}

\usepackage{amsfonts}
\usepackage{amsmath}

\newcommand{\Photo}{}

\begin{document}
\vspace*{2.49cm}
\title{HAMILTONIAN VS STABILITY\\
IN ALTERNATIVE THEORIES OF GRAVITY\,\footnote{Contribution
to the 2019 Gravitation session of the 54th Rencontres de Moriond.}}

\author{G.~ESPOSITO-FARESE}

\address{Sorbonne Universit\'e, CNRS, UMR7095,
Institut d'Astrophysique de Paris,\\
${\mathcal{G}}{\mathbb{R}}\varepsilon{\mathbb{C}}{\mathcal{O}}$,
98bis boulevard Arago, F-75014 Paris, France}

\maketitle\abstracts{
When a Hamiltonian density is bounded by below, we know that the
lowest-energy state must be stable. One is often tempted to
reverse the theorem and therefore believe that an unbounded
Hamiltonian density always implies an instability. The main
purpose of this presentation (which summarizes my
work\,\cite{Babichev:2017lmw,Babichev:2018uiw} with E.~Babichev,
C.~Charmousis and A.~Leh\'ebel) is to pedagogically explain why
this is erroneous. Stability is indeed a coordinate-independent
property, whereas the Hamiltonian density does depend on the
choice of coordinates. In alternative theories of gravity, like
k-essence or Horndeski theories, the correct stability criterion
is a subtler version of the well-known ``Weak Energy Condition''
of general relativity. As an illustration, this criterion is
applied to an exact Schwarzschild-de Sitter solution of a
Horndeski theory, which is found to be stable for a given range
of its parameters, contrary to a claim in the literature.}

\section{Introduction}\label{sec1}
It is well-known that any logical implication such as
$(A\Rightarrow B)$ may also be rewritten as its contrapositive
$(\neg B \Rightarrow \neg A)$, but that its \textit{inverse}
$(\neg A \Rightarrow \neg B)$ is generally not true ---unless the
initial implication was actually an equivalence. It happens that
several recent works in theoretical physics have used such an
erroneous inversion of a standard theorem about stability. The
correct theorem tells us that if a Hamiltonian density is bounded
by below, then the lowest-energy state is stable. Indeed, imagine
a Hamiltonian having the shape of a U: If the solution
corresponds to its minimum, it is impossible to move away without
violating the conservation of energy. One may be tempted to
inverse the statement by claiming that ``a Hamiltonian density
which is unbounded from below always imply an instability''. We
shall underline why this is erroneous, although many physicists
intuitively believe so, probably guided by their experience with
ghost degrees of freedom. Of course, not all solutions are
stable, otherwise the above standard theorem would not have any
interest. In particular, ghosts do lead to deadly instabilities.
But we shall see that stable solutions may sometimes correspond
to a Hamiltonian density which is unbounded from below. Actually,
this is rather trivial once it is understood, but we believe it
is important to be underlined, as this changes some conclusions
of recent papers about the stability of specific solutions. Let
us for instance mention a series of technically excellent
works,\cite{Ogawa:2015pea,Takahashi:2015pad,%
Takahashi:2016dnv,Kase:2018voo} whose calculations are highly
non-trivial and correct, but which unfortunately use the above
erroneous ``inverse'' argument. They compute the Hamiltonian of
perturbations around a given background solution, and impose it
to be bounded by below. Of course, all the stable cases that they
report are indeed (perturbatively) stable, because of the
standard theorem recalled above. But they may be discarding other
stable cases, which do not satisfy their requirement of a
bounded-by-below Hamiltonian. As an example, the first of these
references\,\cite{Ogawa:2015pea} claims that a hairy black-hole
solution of a given Horndeski theory is \textit{always} unstable,
whereas we shall see in Sec.~\ref{sec3} below that it is stable
for a given range of the theory
parameters.\cite{Babichev:2017lmw,Babichev:2018uiw}
Let us start in Sec.~\ref{sec2} with a discussion of the
indirect relation between Hamiltonian and stability.

\section{Stability in presence of several causal cones}\label{sec2}
\subsection{Causal cones and Hamiltonian}\label{subsec21}
\begin{figure}
\centerline{\includegraphics[width=\linewidth]{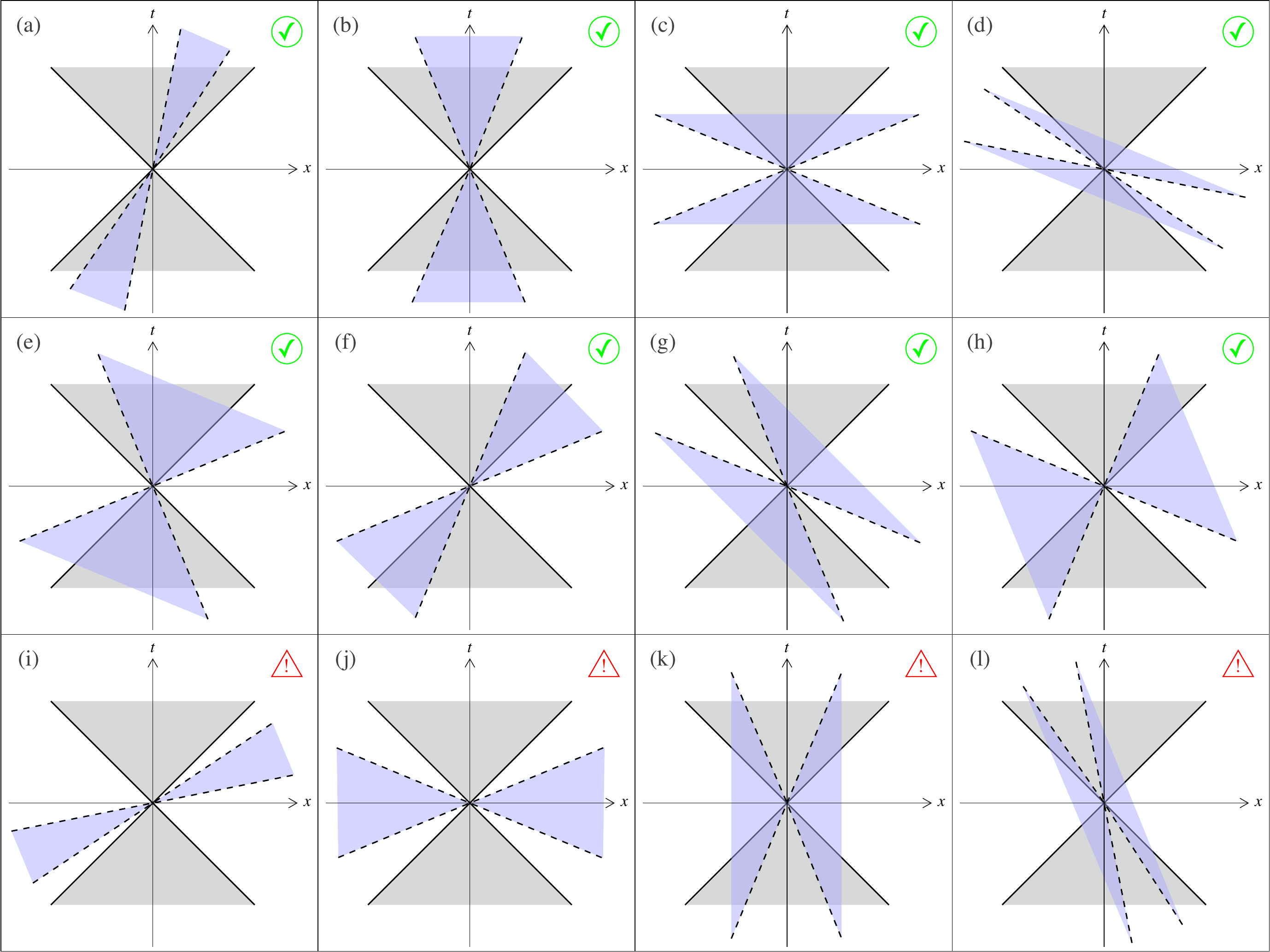}}
\caption[]{Possible relative orientations of two causal cones,
in a coordinate system such that the grey cone with solid lines
appears at $\pm 45^\circ$. We do not plot the equivalent
configurations exchanging left and right, and do not consider
the limiting cases where some characteristics coincide. In our
discussion of k-essence in Sec.~\ref{subsec21}, we assume that
the grey cone is defined by $g^{\mu\nu}$, while the dashed (blue)
one is defined by the effective metric $\mathcal{G}^{\mu\nu}$ in
which spin-0 degrees of freedom propagate.}
\label{Fig:CausalCones}
\end{figure}

Although our discussion is quite general, as this will become
clear after having understood it, let us focus for simplicity on
\textit{perturbative} stability around a given background, and
also only on kinetic terms, as this is where subtleties are
hidden. In alternative theories of gravity, there generically
exist different causal cones for the various propagating degrees
of freedom. The simplest example is k-essence, whose Lagrangian
reads $\mathcal{L} = -\frac{1}{2}f(X)$, where $X\equiv
g^{\mu\nu}\partial_\mu\varphi\partial_\nu\varphi$ would be the
standard kinetic term for a scalar field $\varphi$, and $f$ is
a function specifying the theory. [We choose the mostly-plus
signature convention for the metric $g_{\mu\nu}$.] Let us write
$\varphi = \bar\varphi+\chi$, where $\bar\varphi$ denotes the
background solution and $\chi$ a small perturbation.
Then the second-order expansion of the Lagrangian reads
$\mathcal{L}_2 =-\mathcal{G}^{\mu\nu}
\partial_\mu\chi \partial_\nu\chi$,
where\,\cite{Aharonov:1969vu,ArmendarizPicon:1999rj,%
Babichev:2006vx,Bruneton:2006gf,Bruneton:2007si,Babichev:2007dw}
\begin{equation}
\mathcal{G}^{\mu\nu} = f'(\bar X) g^{\mu\nu} + 2 f''(\bar X)
\nabla^\mu\bar\varphi \nabla^\nu\bar\varphi
\label{Eq:GmunuKessence}
\end{equation}
depends on the first and second derivatives of function $f$
with respect to its argument $\bar X$, and plays the role of an
effective metric in which the spin-0 degree of freedom $\chi$
propagates. When the background gradient $\nabla^\mu\bar\varphi$
does not vanish, the last term of Eq.~\ref{Eq:GmunuKessence}
implies that null directions with respect to
$\mathcal{G}^{\mu\nu}$ are not null with respect to $g^{\mu\nu}$
and reciprocally. This corresponds to panel (b) of
Fig.~\ref{Fig:CausalCones} when $f''<0$, or to panel
(c) when $f''>0$.

The difficulties arise when one performs a boost with a large
enough velocity: Panel (b) is transformed into (a), in which the
time axis gets out of the scalar (dashed blue) causal cone, and
panel (c) becomes (d), where the spatial $x$ axis enters the
scalar cone, on the contrary.\footnote{Note that although one
characteristic goes backwards in time in panel (d) of
Fig.~\ref{Fig:CausalCones}, causality is anyway preserved,
as was discussed in detail more than a decade
ago.\cite{Babichev:2006vx,Bruneton:2006gf,Bruneton:2007si,%
Babichev:2007dw} If the scalar cone never totally opens, i.e.,
that its exterior exists everywhere, then one may foliate the
full spacetime with hypersurfaces which are spacelike with
respect to both $\mathcal{G}^{\mu\nu}$ and $g^{\mu\nu}$, and it
is impossible to influence one's past without assuming a
non-trivial topology. The only subtlety with panel (d) is that
one is not allowed to use the $t=0$ hypersurface as a Cauchy
surface, since it is not spacelike with respect to the scalar
causal cone. The data for the scalar perturbation $\chi$ are thus
obviously constrained on this $t=0$ hypersurface.} This causes
the Hamiltonian density
\begin{equation}
\mathcal{H}_2 =
-\mathcal{G}^{00} \dot\chi^2
+\mathcal{G}^{ij} \partial_i\chi \partial_j\chi
\label{Eq:H2}
\end{equation}
not to be bounded by below any longer. Indeed, let us denote as
$\mathcal{G}_{\mu\nu}$ the \textit{inverse} of the effective
metric $\mathcal{G}^{\mu\nu}$ (beware not to confuse it with
$g_{\mu\lambda}g_{\nu\rho}\mathcal{G}^{\lambda\rho}$). In the
$(t,x)$ subspace of Fig.~\ref{Fig:CausalCones}, it reads
\begin{equation}
\begin{pmatrix}
\mathcal{G}_{00} & \mathcal{G}_{0x} \\
\mathcal{G}_{0x} & \mathcal{G}_{xx}
\end{pmatrix}
= \begin{pmatrix}
\mathcal{G}^{xx} & -\mathcal{G}^{0x} \\
-\mathcal{G}^{0x} & \mathcal{G}^{00}
\end{pmatrix}/D,
\label{Eq:CovMetric}
\end{equation}
where the determinant $D\equiv \mathcal{G}^{00}\mathcal{G}^{xx} -
\left(\mathcal{G}^{0x}\right)^2$ must be strictly negative for
this effective metric to define a cone (and thereby hyperbolic
field equations). In the situation of panel (a) of
Fig.~\ref{Fig:CausalCones}, the time coordinate is spacelike with
respect to the scalar (dashed blue) causal cone, therefore
$\mathcal{G}_{00}\, dx^0\, dx^0 > 0$. Because of
Eq.~\ref{Eq:CovMetric}, this implies $\mathcal{G}^{xx} < 0$.
Similarly, in the situation of panel~(d), the spatial $x$ axis is
timelike with respect to the scalar causal cone, therefore
$\mathcal{G}_{xx}\, dx\, dx < 0$, which implies $\mathcal{G}^{00}
> 0$ from Eq.~\ref{Eq:CovMetric}. In both cases, we thus find
that the Hamiltonian density, Eq.~\ref{Eq:H2}, contains
a term, either $\mathcal{G}^{xx}(\partial_x \chi)^2$ or
$-\mathcal{G}^{00} \dot \chi^2$, which can become infinitely
negative.

On the other hand, the Hamiltonian density, Eq.~\ref{Eq:H2}, is
positive in situations corresponding to panels (b) or (c),
because the time axis is timelike and the $x$ axis spacelike with
respect to the scalar causal cone, therefore $\mathcal{G}^{xx} >
0$ and $\mathcal{G}^{00} < 0$. This implies that the solution is
perturbatively stable, because of the standard theorem recalled
at the beginning of the Introduction: It cannot decay toward
another state without violating energy conservation. Since panels
(a) and (d) correspond to strictly the \textit{same} solutions as
(b) and (c), merely seen by a moving observer, they must
therefore also describe stable cases, in spite of the
unboundedness by below of the Hamiltonian.

Before explaining the deep reason why such cases are indeed
stable, in Sec.~\ref{subsec22} below, we may already draw
conclusions from our argument above: If there exists a coordinate
system in which the Hamiltonian density is bounded by below, then
the standard theorem implies that the solution is stable ---even
if it is unbounded by below in other coordinate systems. In
terms of the spacetime diagrams of Fig.~\ref{Fig:CausalCones},
stability means thus that all causal cones should have a common
interior (intersection of the grey and blue cones), where a new
time axis may be chosen, and also a common exterior (white region
in Fig.~\ref{Fig:CausalCones}), where a new spatial $x$ axis may
be defined. As discussed above, panels (a) and (b) are thus
equivalent, as well as (c) and (d), and they all describe stable
cases.

The same reasoning shows that all four panels (e)--(h) also
correspond to stable solutions: There always exists a coordinate
system (not necessarily obtained by a mere boost) bringing them
to the case of panel (e), where the time axis is timelike with
respect to all causal cones, and the $x$ axis is spacelike, so
that the Hamiltonian density, Eq.~\ref{Eq:H2}, becomes positive.
This second row of Fig.~\ref{Fig:CausalCones} is interesting for
two reasons. First, it illustrates the unusual situation in which
the two metrics $g^{\mu\nu}$ and $\mathcal{G}^{\mu\nu}$ cannot be
simultaneously diagonalized. We know that two quadratic forms can
always be simultaneously diagonalized when at least one of them
is positive (or negative) definite, but here both metrics have a
hyperbolic signature. In this second row, one of the scalar's
characteristics is within the grey cone, and the other one
outside it, therefore no coordinate transformation can bring them
to a symmetric configuration like those of panels (b), (c), (j)
or (k). The second instructive point is about panel (h): The time
axis is spacelike with respect to the scalar causal cone, while
the $x$ axis is timelike with respect to this blue cone.
Therefore, the contribution of the scalar degree of freedom to
the Hamiltonian density, Eq.~\ref{Eq:H2}, is \textit{always}
negative in this coordinate system. On the other hand, usual
matter fields minimally coupled to $g_{\mu\nu}$ give a positive
contribution to the Hamiltonian density, since the grey causal
cone has a standard orientation with respect to the coordinate
axes. Panel (h) corresponds thus to a situation in which the
Hamiltonian density is the sum of a negative (scalar)
contribution and a positive (matter) one. One may thus be tempted
to naively conclude that this solution should decay into an
infinite number of negative-energy scalar modes, compensated by
an infinite number of positive-energy matter ones. However, there
exists another coordinate system, in which the time axis is
chosen in the intersection of the grey and blue cones, and the
$x$ axis in the white region outside both of them, such that the
total Hamiltonian (of matter plus the scalar field) is bounded by
below. The standard theorem therefore implies that this solution
must be stable, and we shall better understand why in
Sec.~\ref{subsec22}.

The last row of Fig.~\ref{Fig:CausalCones}, panels (i)--(l),
describes the unstable cases. The two metrics can be
simultaneously diagonalized by a coordinate change, transforming
panel (i) into (j) and (l) into (k). But in the $(t,x)$ subspace
of this figure, panels (j) and (k) correspond to
\textit{opposite} signatures of the two metrics $g^{\mu\nu}$ and
$\mathcal{G}^{\mu\nu}$: One is $(-,+)$ and the second $(+,-)$.
Therefore, the scalar degree of freedom behaves as a ghost in
this $(t,x)$ subspace, and the solution will indeed decay into an
infinite amount of negative-energy scalar modes compensated by
positive-energy matter ones. For this last row of
Fig.~\ref{Fig:CausalCones}, there does not exist \textit{any}
coordinate system in which the total Hamiltonian density (of
matter plus the scalar field) is bounded by below.

\subsection{Conserved quantities}\label{subsec22}
Let us now explain why some solutions may be stable in spite of
their unbounded Hamiltonian density ---including in the worst
case of panel (h). The reason is that energy is not the only
conserved quantity. Since Lagrangian $\mathcal{L}_2$ (defined
above Eq.~\ref{Eq:GmunuKessence}) is a scalar, it is invariant
under time and space translations, therefore there exist four
Noether currents
\begin{equation}
-T_\mu^\nu \equiv
\frac{\delta \mathcal{L}_2}{\delta (\partial_\nu \chi)}\,
\partial_\mu \chi - \delta_\mu^\nu\, \mathcal{L}_2,
\label{Eq:Noether}
\end{equation}
where $\mu$ specifies which current is considered and $\nu$
denotes its components. When integrating their conservation
equation $\partial_0 T^0_\mu + \partial_i T^i_\mu = 0$ (here
written in flat spacetime to simplify) over a large spatial
volume $V$ containing the whole physical system, the spatial
derivatives become vanishing boundary terms, and one gets the
standard conservation laws for total energy and momentum,
$\partial_t P_\mu = 0$, with $P_\mu \equiv
-\int\!\!\!\int\!\!\!\int_V T_\mu^0\, d^3 x$. For $\mu = 0$, the
energy density $-T_0^0$ coincides with Eq.~(\ref{Eq:H2}). As
shown above, even if one starts from a positive value of $-T_0^0$
in a coordinate system corresponding to panels (b), (c) or (e) of
Fig.~\ref{Fig:CausalCones}, the total energy $P'_0 = (\partial
x^\mu/\partial x'^0)P_\mu$ may become negative in another
coordinate system ---corresponding to panels (a), (d), (f), (g)
or (h). But all four quantities $P'_\lambda$ are anyway
conserved, in this new coordinate system, and it happens that
there exists a linear combination of them which is bounded by
below. Indeed, $P_0 = (\partial x'^\lambda/\partial
x^0)P'_\lambda$ gives precisely the positive energy which was
computed in the initial coordinate system of panels (b), (c) or
(e). In conclusion, when the Hamiltonian density is not bounded
by below, there may anyway exist a linear combination of the four
conserved Noether currents which is bounded by below, and its
existence suffices to ensure stability ---for the same reason as
in the standard theorem recalled at the beginning of the
Introduction. This conserved and bounded-by-below quantity
actually coincides with the Hamiltonian computed in a ``good''
coordinate system, such that time is timelike and space spacelike
with respect to all causal cones, consistently with our
conclusions of Sec.~\ref{subsec21}.

\subsection{Stability criterion}\label{subsec23}
The eight stable cases (a)--(h) of Fig.~\ref{Fig:CausalCones} may
be translated as conditions on the components of the effective
metric $\mathcal{G}^{\mu\nu}$. In the $(t,x)$ subspace of this
figure, and in a coordinate system such that $g_{\mu\nu} =
\text{diag}(-1,1)$, one finds that stability requires
\begin{eqnarray}
D\equiv \mathcal{G}^{00}\mathcal{G}^{xx}
- \left(\mathcal{G}^{0x}\right)^2 < 0&& \text{(hyperbolicity),}
\label{Eq:hyperbolicity}\\
\mathcal{G}^{00} < \mathcal{G}^{xx}\quad\text{and/or}\quad
|\mathcal{G}^{00}+\mathcal{G}^{xx}| < 2 |\mathcal{G}^{0x}|
&& \vtop{\hbox{(existence of consistent}
\hbox{time and space coordinates).}}
\label{Eq:consistency}
\end{eqnarray}
This means that the off-diagonal component $\mathcal{G}^{0x}$
should be large enough. For instance, if
\hbox{$0<\mathcal{G}^{00} < \mathcal{G}^{xx}$}, then
Eq.~\ref{Eq:hyperbolicity} implies than $\mathcal{G}^{0x}$ is
large enough to ensure the existence of a coordinate system in
which time is timelike and space spacelike with respect to all
causal cones. But if \hbox{$0< \mathcal{G}^{xx}
<\mathcal{G}^{00}$}, then Eq.~\ref{Eq:consistency} implies
that $|\mathcal{G}^{0x}|$ must be even larger, namely greater
than the arithmetical mean
$\frac{1}{2}|\mathcal{G}^{00}+\mathcal{G}^{xx}|$,
known to be always greater than the geometrical mean
$\sqrt{\mathcal{G}^{00}\mathcal{G}^{xx}}$ entering
Eq.~\ref{Eq:hyperbolicity}. By contrast, the positivity of the
Hamiltonian, Eq.~\ref{Eq:H2}, would need $\mathcal{G}^{00} < 0$
and $\mathcal{G}^{xx} > 0$, which is much more restrictive than
Eqs.~\ref{Eq:hyperbolicity}-\ref{Eq:consistency} above, and does
not depend at all on $\mathcal{G}^{0x}$. This shows that some
stable solutions may have been wrongly discarded in the recent
literature.\cite{Ogawa:2015pea,Takahashi:2015pad,%
Takahashi:2016dnv,Kase:2018voo} Actually, some of these
references chose to replace the positivity of the Hamiltonian by
the ``necessary'' condition $\mathcal{G}^{00}\mathcal{G}^{xx}
< 0$. But contrary to the hyperbolicity condition,
Eq.~\ref{Eq:hyperbolicity}, this inequality is obviously
coordinate-dependent: As illustrated in Sec.~\ref{subsec21},
different observers may find opposite signs for the product
$\mathcal{G}^{00}\mathcal{G}^{xx}$, whereas stability is a
physical statement which should be coordinate-independent.

To simplify, the second line of the above stability conditions,
Eq.~\ref{Eq:consistency}, has been written in a specific
coordinate system such that $g_{\mu\nu} = \text{diag}(-1,1)$. It
may of course be generalized to an arbitrary coordinate system,
but it is more useful to express it in a covariant way. We
found\,\cite{Babichev:2018uiw} that the necessary and sufficient
conditions for stability are the following. First, all metrics
(here $g_{\mu\nu}$ and $\mathcal{G}_{\mu\nu}$, but there may
exist more for other degrees of freedom) should be of hyperbolic
mostly-plus signature. This generalizes
Eq.~\ref{Eq:hyperbolicity} above. Second, there should exist at
least one contravariant vector $U^\mu$ and one covariant vector
$u_\mu$ (generically not related to each other by raising or
lowering their index with any of the metrics) such that
\begin{eqnarray}
g_{\mu\nu} U^\mu U^\nu < 0, &\quad&
\mathcal{G}_{\mu\nu} U^\mu U^\nu < 0, \quad \dots
\label{Eq:gUU}\\
g^{\mu\nu} u_\mu u_\nu < 0, &\quad&
\mathcal{G}^{\mu\nu} u_\mu u_\nu < 0, \hskip 1.25pc \dots
\label{Eq:guu}
\end{eqnarray}
for all metrics (where we recall that $\mathcal{G}_{\mu\nu}$
denotes the \textit{inverse} of $\mathcal{G}^{\mu\nu}$).
Equation~\ref{Eq:gUU} implies the existence of a common interior
to all causal cones, where a ``good'' time axis may be chosen,
namely $dx^0$ in the direction of $U^\mu$. Equation~\ref{Eq:guu}
expresses the existence of a spatial hypersurface exterior to all
causal cones, defined by $u_\mu dx^\mu = 0$, where ``good''
spatial coordinates may be chosen. Finally, the positivity of
the Hamiltonian density in such a good coordinate system may
be covariantly written as
\begin{equation}
T_\mu^\nu U^\mu u_\nu \geq 0,
\label{Eq:positiveT00}
\end{equation}
where $T_\mu^\nu$ denotes the total energy-momentum tensor
for all fields.

Equations~\ref{Eq:gUU}--\ref{Eq:positiveT00} actually generalize
the ``Weak Energy Condition'' of general relativity. When there
exists only one metric $g_{\mu\nu}$ to which all fields are
minimally coupled, and thereby a single causal cone for all
degrees of freedom, one may of course choose $u_\mu = g_{\mu\nu}
U^\nu$, and Eq.~\ref{Eq:positiveT00} becomes the standard
condition $T_{\mu\nu} U^\mu U^\nu \geq 0$ for any timelike vector
$U^\mu$. Note that even in general relativity, the single causal
cone defined by $g_{\mu\nu}$ may be oriented like the dashed blue
cones of Fig.~\ref{Fig:CausalCones}. For instance, in the
interior of a Schwarzschild black hole, Schwarzschild coordinates
define the time axis outside the causal cone ---similarly to the
blue cone of panels (j) or (k). More generally, \textit{any}
coordinate system is allowed in general relativity, even if the
time axis is outside the causal cone and/or some spatial axes
within it. In such cases, it is well-known that the Hamiltonian
density $-T_0^0$ is not positive, but that the correct stability
criterion involves the contraction $T_{\mu\nu} U^\mu U^\nu$ with
a timelike vector $U^\mu$ (i.e., such that $g_{\mu\nu} U^\mu
U^\nu < 0$). Our conclusions above are straightforward
generalizations: One should never trust coordinate-dependent
reasonings, and the Hamiltonian density does depend on the
coordinate system, since it is not a scalar quantity.

As an example, let us quote the theoretical constraints imposed
by the above stability conditions,
Eqs.~\ref{Eq:hyperbolicity}-\ref{Eq:consistency} or
Eqs.~\ref{Eq:gUU}--\ref{Eq:positiveT00}, on k-essence theories
(defined above Eq.~\ref{Eq:GmunuKessence}). One recovers those
which have been derived several times in the literature with
various viewpoints.\cite{Aharonov:1969vu,ArmendarizPicon:1999rj,%
Babichev:2006vx,Bruneton:2006gf,Bruneton:2007si,Babichev:2007dw}
One needs $f'(X) > 0$ and $2 X f''(X) + f'(X) > 0$, whatever
the direction of the gradient $\nabla^\mu\varphi$ (timelike,
spacelike or null). This imposes in particular that there exists
a common spacelike exterior to both causal cones (defined by
$g_{\mu\nu}$ and $\mathcal{G}_{\mu\nu}$), where one may specify
initial data. Note that there is no constraint on $f''(X)$ alone,
and therefore that both infraluminal [panels (a) or (b) of
Fig.~\ref{Fig:CausalCones}] and superluminal cases [panels (c)
or (d)] are allowed.

\section{Stable black hole in a Horndeski theory}\label{sec3}
A better illustration of the above stability criterion is
provided by a simple Horndeski theory,\cite{Horndeski:1974wa}
defined by the action
\begin{equation}
S = \int{\left[\zeta (R-2\Lambda_\text{bare})
- \eta \left(\partial_\mu\varphi\right)^2
+\beta\, G^{\mu \nu} \partial_{\mu}\varphi \partial_{\nu}\varphi
\right] \sqrt{-g}\,d^4x},
\label{Eq:Action}
\end{equation}
where $R$ is the scalar curvature of the metric $g_{\mu\nu}$
(to which all matter fields are assumed to be minimally coupled),
$G^{\mu\nu}$ is its Einstein tensor (not to be confused with the
effective metric $\mathcal{G}^{\mu\nu}$ of
Eq.~\ref{Eq:GmunuKessence}), $\Lambda_\text{bare}$ denotes a bare
cosmological constant, and $\zeta$, $\eta$, $\beta$ are constant
parameters. This theory admits an exact Schwarzschild-de Sitter
solution of the form\,\cite{Babichev:2013cya}
\begin{align}
ds^2 &= -A(r)\, dt^2+\dfrac{dr^2}{A(r)}
+ r^2\left(d\theta^2 +\sin^2\theta\, d\phi^2\right),
\label{Eq:ds2}\\
A(r) &=1- \frac{2Gm}{r} - \frac{\Lambda_\text{eff}}{3}\, r^2,
\hskip 3.25pc\text{with }~\Lambda_\text{eff} = -\frac{\eta}{\beta},
\label{Eq:A}\\
\varphi &= q\left[ t - \int\frac{\sqrt{1-A(r)}}{A(r)}\, dr\right],
\quad\text{with }~q^2 = \frac{\eta+
\beta\, \Lambda_\text{bare}}{\eta\,\beta}\,\zeta.
\label{Eq:phi}
\end{align}
Its interesting property is that the observable cosmological
constant $\Lambda_\text{eff}$, entering the line element $ds^2$
through Eq.~\ref{Eq:A}, is not the bare one of action
Eq.~\ref{Eq:Action}, but an effective one which may be small
enough to be consistent with observation even if
$\Lambda_\text{bare}$ is huge (for instance the square of the
Planck mass, or even larger). In the present model,
$\Lambda_\text{eff}$ does not even depend at all on
$\Lambda_\text{bare}$, but only on the two kinetic terms defining
the dynamics of the scalar field $\varphi$ in
Eq.~\ref{Eq:Action}. This is a particularly nice example of what
is called ``self-tuning'':\,\footnote{It was later
shown\,\cite{Babichev:2016kdt} that such a self-tuning can be
achieved in basically \textit{all} Horndeski and beyond-Horndeski
theories, provided the action contains at least two of the six
possible terms defining the scalar's dynamics.} The scalar field
automatically adjusts itself so that its energy-momentum tensor
$T_{\mu\nu}$ almost perfectly balances the vacuum energy
$\Lambda_\text{bare}\, g_{\mu\nu}$ entering Einstein's equations,
in order to let a tiny observable one $\Lambda_\text{eff}\,
g_{\mu\nu}$.

To analyze the stability of such a solution, we need to extract
the effective metrics in which spin-0 and spin-2 perturbations
propagate. The most efficient method would be to find a change of
variables diagonalizing their kinetic terms, i.e., what is called
the ``Einstein frame''. The procedure is well-known for standard
(Jordan-Fierz-Brans-Dicke) scalar-tensor theories or $f(R)$
theories, and this was also achieved for the quadratic plus cubic
Galileon model,\cite{Babichev:2012re} but we did not find any
covariant change of variables separating the degrees of freedom
in the present theory. We have thus studied perturbations by
decomposing them on spherical
harmonics.\,\cite{Ogawa:2015pea,Babichev:2017lmw,Babichev:2018uiw}

\begin{figure}
\centerline{\includegraphics[width=\linewidth]{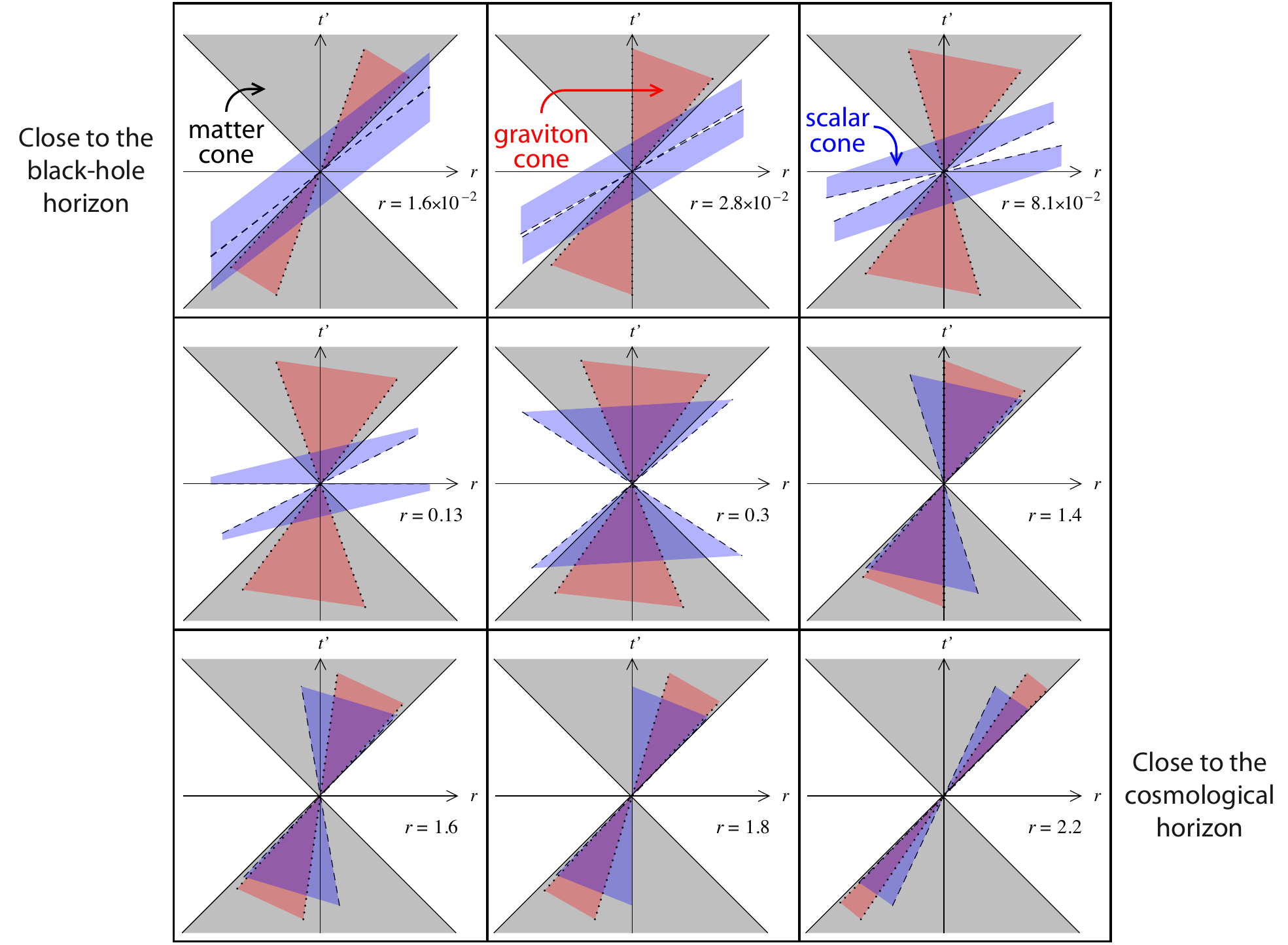}}
\caption{Matter (solid grey), graviton (dotted red) and scalar
(dashed blue) causal cones in the exact Schwarzschild-de Sitter
solution of Eqs.~\ref{Eq:ds2}--\ref{Eq:phi}, for parameters
$\beta=-1$ and $\zeta = \Lambda_\text{bare} = 2\eta = 1$ in
Planck units. Static Schwarzschild coordinates are used, but the
time axis has been rescaled so that the grey cone (defined by
$g_{\mu\nu}$) appears at $\pm 45^\circ$. The successive panels
correspond to situations from close to the black-hole horizon
at the top-left, to close to the cosmological horizon at the
bottom-right.}
\label{Fig:bhsoundcone}
\end{figure}

Odd-parity perturbations necessarily correspond to the spin-2
degree of freedom, and we fully agree with the analytical results
previously derived in the literature.\cite{Ogawa:2015pea}
However, the conclusion of this reference was that this solution
is always unstable, because the product
$\mathcal{G}^{00}\mathcal{G}^{rr}$ becomes positive close enough
to the black-hole horizon (see our discussion below
Eq.~\ref{Eq:consistency}). Instead of writing a heavy analytical
expression for this product, let us plot the causal cones in
Fig.~\ref{Fig:bhsoundcone}.
The dotted red ones represent the graviton causal cone we are
presently discussing. The fact that
$\mathcal{G}^{00}\mathcal{G}^{rr}$ becomes positive is vividly
illustrated by the top-left panel: The time axis gets out of this
red cone. However, it is also clear that one may choose another
time axis within this graviton causal cone (which happens to be
itself inside the matter causal cone\,\footnote{See our published
articles\,\cite{Babichev:2017lmw,Babichev:2018uiw} for a
modification (in the beyond-Horndeski class of theories) of the
model of the present Sec.~\ref{sec3}, in which the matter and
graviton causal cones exactly coincide everywhere.}), and the
perturbation Hamiltonian will become positive in this new
coordinate system. Therefore, there is actually no instability
caused by the gravitons. It is also easy to check on
Fig.~\ref{Fig:bhsoundcone} that at any distance from the black
hole, there always exist a common interior and a common exterior
to the matter and graviton causal cones. Actually, note that the
graviton cone is also very tilted near the cosmological horizon
(bottom-right panel). Therefore, the same argument as the
literature\,\cite{Ogawa:2015pea} about the sign of
$\mathcal{G}^{00}\mathcal{G}^{rr}$ would have also concluded that
the solution must be unstable. But this is again a coordinate
artifact, caused here by the static Schwarzschild coordinates
used in this figure. If one used
Friedmann-Lema\^{\i}tre-Robertson-Walker coordinates instead, the
red (graviton) cone would remain thinner than the grey (matter)
one, close to the cosmological horizon, but it would be perfectly
centered and the time axis would be inside it.

To prove stability, we also need to study spin-0 perturbations.
In order to extract the effective metric in which they propagate,
we focused on the $\ell = 0$ (spherically symmetric) even-parity
modes, which can only describe a scalar degree of freedom. We
found\,\cite{Babichev:2017lmw,Babichev:2018uiw} that the scalar
cone has a consistent orientation with the matter and graviton
cones (i.e., both a common interior and a common exterior) if and
only if
\begin{align}
\mathrm{either}~\eta&>0,~\beta
<0\quad\text{and}\quad\dfrac{\Lambda_\text{bare}}{3}
<-\dfrac{\eta}{\beta}<\Lambda_\text{bare},
\label{Eq:range1}
\\
\mathrm{or}~\eta&<0,~\beta>0\quad\text{and}\quad\Lambda_\text{bare}
<-\dfrac{\eta}{\beta}<3\Lambda_\text{bare}.
\label{Eq:range2}
\end{align}
Since Eq.~\ref{Eq:A} tells us that $\Lambda_\text{eff} =
-\eta/\beta$, these conditions actually prove that self-tuning is
impossible in the model of Eq.~\ref{Eq:Action}: The observed
cosmological constant $\Lambda_\text{eff}$ can never be
negligible with respect to the bare one $\Lambda_\text{bare}$,
otherwise the solution is unstable. However, this model and its
solution are experimentally viable if $\Lambda_\text{bare}$ is
assumed to be small enough, like in general relativity, and
Fig.~\ref{Fig:bhsoundcone} illustrates that it is stable when
Eq.~\ref{Eq:range1} is satisfied. Indeed the scalar (dashed blue)
causal cone remains everywhere consistent with the matter and
graviton causal cones. It is difficult to see what happens near
the black-hole horizon, because this scalar causal cone opens
almost totally. This is again an coordinate artifact, because we
chose to plot the matter (grey) cone at $\pm 45^\circ$. This
matter cone actually becomes infinitely thin near the black-hole
horizon, in Schwarzschild coordinates, and our rescaling of the
time coordinate in Fig.~\ref{Fig:bhsoundcone} is thus responsible
for the wide opening of the scalar (blue) cone. But its exterior
always exists, and there is thus always a common exterior to all
three causal cones, where one may specify initial data.

\section{Conclusions}\label{sec4}
The main message of this presentation is that a Hamiltonian
density which is unbounded from below does not always imply an
instability. Indeed, the 3-momentum is also conserved, and it may
be linearly combined with the energy to give a bounded-by-below
quantity ---which actually coincides with the Hamiltonian
computed in another coordinate system. The simplest way to
analyze the stability of a solution is to plot the causal cones
of all degrees of freedom: There should exist both a common
interior and a common exterior spacelike hypersurface. This
stability criterion may be expressed as a generalization of the
Weak Energy Condition of general relativity, now encompassing
the case of several causal cones. Finally, we illustrated this
criterion by showing that an exact Schwarzschild-de Sitter
solution of a Horndeski theory is stable for a given range
of its parameters, contrary to a claim in the literature.

\end{document}